\documentclass[sigconf,table,dvipsnames,xcdraw,screen]{acmart}
\usepackage{courier}
\usepackage[ruled,vlined]{algorithm2e}
\usepackage{graphicx}
\usepackage{paralist}
\usepackage[shortlabels]{enumitem}
\usepackage{tabularx}
\usepackage{balance}
\usepackage{multirow}
\usepackage{multicol}
\usepackage{pbox}
\usepackage{mathtools}
\usepackage[TABBOTCAP]{subfigure}
\usepackage{algorithmic}
\usepackage{paralist}
\usepackage{tabularx}
\usepackage{url}
\usepackage{balance}
\usepackage{multirow}
\usepackage{multicol}
\usepackage{setspace}
\usepackage{enumitem}
\usepackage{verbatim}
\usepackage{float}
\usepackage[normalem]{ulem}
\usepackage{xspace}
\usepackage{ifthen}
\usepackage{bold-extra}
\usepackage{hyperref}
\usepackage{pifont}
\usepackage{tikz}
\usepackage{xcolor}
\usepackage{enumitem}
\usepackage{array}
\usepackage{multirow}

\usepackage{booktabs,xcolor}

\newcommand\MyBox[2]{
  \fbox{\lower0.75cm
    \vbox to 1.7cm{\vfil
      \hbox to 1.7cm{\hfil\parbox{1.4cm}{#1\\#2}\hfil}
      \vfil}%
  }%
}

\definecolor{MidnightBlue}{HTML}{01693F}
\include{macro}
\newcommand{\MotorEaseB}{{\bfseries\scshape MotorEase}\xspace}
\newcommand{\MotorEase}{{\sc MotorEase}\xspace}
\newcommand{\MotorCheck}{{\sc MotorCheck}\xspace}

\newcommand{\AidUI}{{\sc AidUI}\xspace}

\copyrightyear{2024}
\acmYear{2024}
\setcopyright{rightsretained}
\acmConference[ICSE '24]{2024 IEEE/ACM 46th International Conference on Software Engineering}{April 14--20, 2024}{Lisbon, Portugal} \acmBooktitle{2024 IEEE/ACM 46th International Conference on Software Engineering (ICSE '24), April 14--20, 2024, Lisbon, Portugal}\acmDOI{10.1145/3597503.3639167}
\acmISBN{979-8-4007-0217-4/24/04}

\begin{document}

\title[\MotorEase: Automated Detection of Motor Impairment\\ Accessibility Issues in Mobile App UIs]{\MotorEaseB: Automated Detection of Motor Impairment Accessibility Issues in Mobile App UIs}

\author{Arun Krishnavajjala}
\email{akrishn@gmu.edu}
\affiliation{%
	\institution{George Mason University}
	\country{Farifax, VA, USA}
}

\author{SM Hasan Mansur}
\email{smansur4@gmu.edu}
\affiliation{%
	\institution{George Mason University}
	\country{Fairfax, VA, USA}
}

\author{Justin Jose}
\email{justinjp2028@gmail.com}
\affiliation{%
	\institution{South Lakes High School}
	\country{Reston, VA, USA}
}

\author{Kevin Moran}
\email{kpmoran@ucf.edu}
\affiliation{%
	\institution{University of Central FL}
	\country{Orlando, FL, USA}
}

\begin{abstract}
Recent research has begun to examine the potential of automatically finding and fixing accessibility issues that manifest in software. However, while recent work makes important progress, it has generally been skewed toward identifying issues that affect users with certain disabilities, such as those with visual or hearing impairments. However there are other groups of users with different types of disabilities that also need software tooling support to improve their experience. As such, this paper aims to automatically identify accessibility issues that affect users with \textit{motor-impairments}. 

To move toward this goal, this paper introduces a novel approach, called \MotorEase, capable of identifying accessibility issues in mobile app UIs that impact \textit{motor-impaired users}. Motor-impaired users often have limited ability to interact with touch-based devices, and instead may make use of a switch or other assistive mechanism --- hence UIs must be designed to support both limited touch gestures and the use of assistive devices. \MotorEase adapts computer vision and text processing techniques to enable a semantic understanding of app UI screens, enabling the detection of violations related to four popular, previously unexplored UI design guidelines that support motor-impaired users, including: (i) visual touch target size, (ii) expanding sections, (iii) persisting elements, and (iv) adjacent icon visual distance. We evaluate \MotorEase on a newly derived benchmark, called \MotorCheck, that contains 555 manually annotated examples of violations to the above accessibility guidelines, across 1599 screens collected from 70 applications via a mobile app testing tool. Our experiments illustrate that \MotorEase is able to identify violations with an average accuracy of $\approx$90\%, and a false positive rate of less than 9\%, outperforming baseline techniques.

\end{abstract}



\maketitle

\vspace{-0.6em}
\section{Introduction}
\label{sec:introduction}

The everyday lives of end-users depend on the use of software applications that support critical tasks such as banking, reading news, and communicating with others. Due to the central role of software in modern society, developers have an obligation to ensure that people of \textit{all} abilities and backgrounds are able to use applications to carry out daily tasks. 
However, this ideal is still very much a goal that engineers must collectively work toward, as past work has illustrated the prevalence of accessibility issues in mobile app software ecosystems~\cite{Alshayban20,Vendome19,Chen22}. Furthermore, the need for accessible software now transcends a moral pursuit, as government agencies worldwide have begun to advocate for more accessible software by introducing legislation which, ``prohibits discrimination on the basis of disability in the activities of public accommodations," ~\cite{ADALaws}.
Beyond providing equitable access to software for users with a variety of backgrounds, accessibility features often improve user experience more broadly, as many accessibility guidelines are designed following the general principals of universal design~\cite{univ-design}, in that the adherence to such guidelines is more likely to lead to an improved user experience for \textit{all} users~\cite{Sarsenbayeva22}.

\begin{figure}[t]
    \centering
    \includegraphics[width=0.4\textwidth]{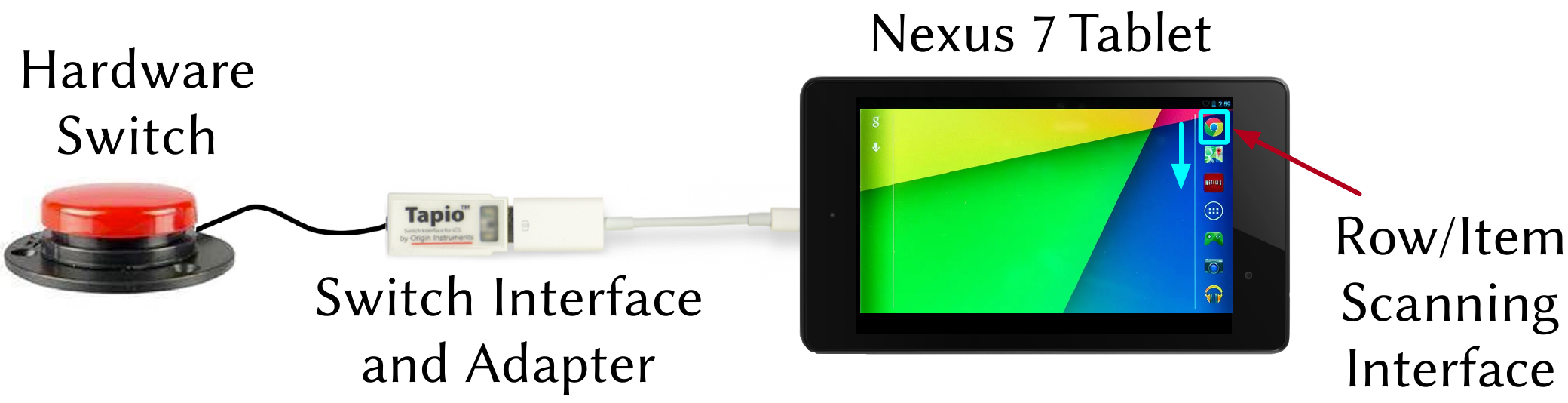}
	\vspace{-1em}
    \caption{Switch Interface}
    \label{fig:switch}
\end{figure}

Current research at the intersection of developer tools and software accessibility has generally been disproportionally focused on users with certain disabilities, such as visual impairments, i.e., low vision (LV), and hearing impairments, i.e., deaf and hard of hearing (DHH)~\cite{Park14,Zhang21,Gajos07,Vendome19, Chen22}. The visual nature of software user interfaces (UIs), and large populations of users with visual impairments have made this a natural and important focus area. This focus, however, must expand to study and create tools that aid developers in considering and implementing accessible features that support users with a wider range of disabilities. One understudied demographic, and the focus of this paper, is that of \textit{motor-impaired users}. The current landscape of research on developer tools that aim to support software accessibility for motor-impaired users is somewhat limited, due in part to the difficulty in supporting a wide spectrum of motor-impairment conditions (i.e., ranging from hand tremors, to more limited motor abilities that necessitate the use of assistive devices such as switch controls) and need to consider external hardware~\cite{Sarsenbayeva22} . 
Generally, developers currently lack tools for identifying, understanding, and implementing accessible features for motor-impaired users~\cite{Alshayban20}. 

The central challenge of building developer tools that support accessibility for motor impaired users is one of \textit{semantic screen understanding}. That is, in order to determine whether a given UI screen follows motor-impairment accessibility guidelines, the \textit{functional} and \textit{visual} properties of UI screens and individual components must be automatically inferred from a given app. For example, motor-impaired users that make use of hardware devices, such as switches, rely on assistive services (e.g., Android \texttt{\small Switch Access}~\cite{switch-access}) that iteratively scan through and highlight individual UI components, as illustrated in Figure~\ref{fig:switch}. This allows a motor-impaired user to easily select the icon or component with which they wish to interact. However, this process can be slower than traditional gesture-based control~\cite{MacKenzie11}, and switch users often rely on the \textit{consistency} of certain UI element patterns, such as menus, in order to quickly perform actions. As such, a popular motor-impairment accessibility design guideline for mobile apps~\cite{AppleAccess,GoogleAccess} states that persistent icons that appear across multiple screens, such as tab bars or menus, should retain a consistent ordering. This allows a user to anticipate which UI elements will be highlighted by the assistive service. However, in order to detect inconsistent icon orderings, an automated tool must be able to accurately identify functional groups of UI components, such as tab bars, and the ordering of icons within them. 

To help advance the current state of developer tools to better support motor impaired users, and overcome the challenges related to automated screen understanding, this paper introduces a novel approach, called \MotorEase, which aims to automate the detection of \textbf{Motor} impairm\textbf{E}nt \textbf{A}cce\textbf{S}sibility issu\textbf{E}s in mobile apps. \MotorEase is a novel approach that leverages automated dynamic analysis, computer-vision, and text-processing techniques to detect violations of motor-impairment accessibility guidelines in a given Android application. The approach is comprised of four detectors each targeted toward a popular UI design guideline meant to support motor impaired users. \MotorEase's novelty lies in both its  technical underpinnings and its ability. \MotorEase combines multiple neural models for screen understanding, allowing it to recognize screen semantics prior techniques cannot. This allows \MotorEase to identify violations of motor-impairment accessibility guidelines. 
\MotorEase is designed to seamlessly integrate into existing software testing workflows, and operates in a fully automatic manner by analyzing common artifacts produced by existing Android testing tools. 
\MotorEase then passes this data to a series of four detectors, each of which identifies guideline violations.
\noindent The main contributions of this paper are as follows:

\begin{enumerate}
	\item \MotorEase ~-- a novel, automated motor-impairment accessibility guideline violation detection tool which is able to accurately detect issues in applications when used in conjunction with an automated inout generation tool;
	\item The \MotorCheck benchmark, consisting of 555 screens that violate one of four popular motor-impairment accessibility guidelines, and 1044 that follow such guidelines, collected from popular Google Play applications.
	\item The results of an empirical evaluation that measures the effectiveness of \MotorEase in detecting motor-impairment accessibility issues within the \MotorCheck benchmark.
	\item A Replication Package~\cite{appendix,site,zenodo} that contains \MotorEase's code, the \MotorCheck benchmark, and our experimental infrastructure to foster replicability and further promote research on motor-impairment accessibility issues.
\end{enumerate}

\vspace{-1em}
\section{Background and Motivation}

\label{sec:motivation}

In order to understand how the \MotorEase approach functions, it is important to understand how motor-impaired users use their devices. Motor-impaired users use devices in two main ways: Switch input and touch input \cite{Zhang13}. As described earlier, switch based input uses an external hardware input device. Switches are common in users with limited to no motor control but without impaired cognitive functions~\cite{Zhang13}. Most current smartphone and tablet devices offer accessibility services that support scanning a curser across the screen to highlight UI elements and icons (\eg Fig.~\ref{fig:switch})~\cite{AppleAccess,GoogleAccess}. The users can then click the switch to select the UI element that is highlighted, forming a semi-automated form of user input. 
This input technique can be far slower than traditional touch-based interactions, as it requires waiting for appropriate UI elements to be highlighted by the system. Thus, switch users rely on consistent and accessible app designs to be able to anticipate the future UI elements that will be focused by the scanning service. One example that highlights how tedious this input can be is related to typing speed, where switch users average 3 words per minute (wpm), and touch-based users average closer to 40 wpm~\cite{MacKenzie11}.

\begin{table*}[h!]
	\centering
	\footnotesize
	\vspace{-0.5em}
	\caption{Accessibility guidelines extracted from our systematic literature review of accessibility guidelines -- includes recent research and Google's~\cite{GoogleAccess} and Apple's~\cite{AppleAccess} accessibility guidelines. (LV = low vision users, DHH = deaf and hard of hearing users)}
	\vspace{-1em}
	\begin{tabular}{>{\centering\arraybackslash}p{2in}|>{\centering\arraybackslash}p{1.8in}|>{\centering\arraybackslash}p{.82in}|>{\centering\arraybackslash}p{.82in}|>{\centering\arraybackslash}p{.82in} }
	
		\textbf{Accessibility Guideline} & \textbf{Primary Affected User Demographic} & \textbf{Guideline Source}  & \textbf{Previous Implementation} & \textbf{Implemented by \MotorEase} \\
		\hline
		Visual Touch Target Size & \footnotesize {Motor, LV} & \cite{Kong21, Parhi06} &  & \cmark \\ 
		\rowcolor{gray!30!} Touch Target Size & \footnotesize {Motor, LV} & \cite{AppleAccess, GoogleAccess, HarvardAccess, WebGuide, Nunes15, Calvo16, Alshayban20, Abascal11, Kane11, Kong21} & \cmark & \\ 
		Persistent Element Location & \footnotesize {Motor} &\cite{AppleAccess, HarvardAccess, WebGuide, GoogleAccess} &   & \cmark \\
		\rowcolor{gray!30!} Clickable Span & \footnotesize {Motor, LV} & \cite{Alshayban20} & \cmark{} &  \\
		Duplicate Clickable Bounds & \footnotesize {Motor} & \cite{Alshayban20} & \cmark{} &  \\
		\rowcolor{gray!30!} Editable Item Descriptions & \footnotesize {LV} & \cite{Alshayban20, Eler18} & \cmark{} &  \\
		Expanding Section Closure & \footnotesize {Motor, LV} & \cite{AppleAccess, GoogleAccess, HarvardAccess, WebGuide} &   & \cmark\\
		\rowcolor{gray!30!} Non-Native Elements & \footnotesize {Motor, LV} & \cite{GoogleAccess, Calvo16}& \cmark{} &  \\
		Motion Activation & \footnotesize {Motor, LV} & \cite{AppleAccess, GoogleAccess, HarvardAccess} &  & \\
		\rowcolor{gray!30!} Labeled Elements & \footnotesize {Visual, DHH} & \cite{Alshayban20, FlrezAristizbal19, Li21, Eler18} & \cmark & \\
		Screen Captioning & \footnotesize {LV, DHH} & \cite{AppleAccess, GoogleAccess, HarvardAccess, WebGuide, ADAWeb, AccessGov, Ross18, Li21, Pavel20, Kane11}& \cmark & \\
		\rowcolor{gray!30!}Keyboard Navigation & \footnotesize {Motor, LV} & \cite{ADAWeb, AccessGov, FlrezAristizbal19, Li21, Chiou21} & \cmark & \\
		Traversal Order & \footnotesize {Motor} & \cite{AccessGov, Alshayban20, FlrezAristizbal19} & \cmark & \\
		\rowcolor{gray!30!}Adjacent Visual Icon Distance & \footnotesize {Motor} & \cite{AppleAccess, GoogleAccess, WebGuide, Yan19, Abascal11, Nunes15} &  & \cmark \\
		
		 Proper Information Organization & \footnotesize {Motor, LV} & \cite{Calvo16} & \cmark & \\
		\rowcolor{gray!30!} Facial Recognition & \footnotesize {Motor} & \cite{Calvo16, Astler11} & \cmark & \\
		 Single Tap Navigation & \footnotesize {Motor} & \cite{AppleAccess, GoogleAccess, HarvardAccess, WebGuide, FlrezAristizbal19, Milne18} &  & \\
		\rowcolor{gray!30!} Poor form design/instructions & \footnotesize {Motor, LV} & \cite{ADAWeb, AccessGov} &  & \\

			\end{tabular}
\vspace{-1em}
	\label{tab:guidelines}
\end{table*}

For users with some, but limited motor-control, touch input is typically used with constraints related to physical limitations such as tremors or gesture speed limitations~\cite{Montague14}. Some users with lesser motor-impairments may also turn to using voice commands to navigate applications due to difficulties/slowness related to switch operation~\cite{Zhang13}. Prior work in the HCI research community has aimed to improve gesture recognition for users with limited motor control~\cite{Peng19}. Google and Apple also offer sensitivity and touch settings that aim to allow users with limited motor control to customize touch controls for easier interaction~\cite{Peng19,AppleAccess,GoogleAccess}. Google and Apple \cite{AppleAccess,GoogleAccess} have also developed UI and human interaction guidelines which aim to support motor-impaired users, but as past work has illustrated, developers may often be unaware of such guidelines or ignore them due to other development constraints~\cite{Salehnamadi21}.

\subsection{Accessibility Guideline Literature Review}

\revision{In order to fully capture the current landscape of accessibility guidelines that may impact various populations of users, and to aid in selecting the most impactful guidelines that aim to assist motor impaired users} we conducted a systematic literature reviews on research at the intersection of software engineering, human-computer interaction, and accessibility. To conduct this review, we followed the methodology set forth by Kitchenham~\etal~\cite{kitchenham2007guidelines}. We defined a single research question that asked \textit{``What accessibility guidelines have been identified and discussed in prior research?''}. We used the relatively simple search string of "accessibility" to search DBLP, the ACM Digital Library, and IEEE Xplore, for work at the intersection of accessibility and software engineering for the date range of January 2010 - December 2022. The purpose of using such a simple search string was to "cast a wide net" and ensure that we did not miss important work. We defined inclusion criteria as follows: (i) must have been published in our studied date range, (ii) must have been published at one of 16 conference venues (ICSE, FSE, ASE, ICSME, MSR, ICPC, ISSTA, ICST, SANER, UIST, CHI, SPLASH, OOPSLA, PLDI, CSCW, ASSETS) or 5 journal venues (TSE, TOSEM, EMSE, JSS, ASE) that cross cut software engineering, HCI, and accessibility, (iii) the paper must describe a study or developer tool directly related to an accessibility issue that impacts end-users. The scope of our search was limited to these venues and digital libraries as they provide the highest quality of research in all matters including accessibility. Our search results returned 2948 papers from our selected conferences within our given date range. Then, two authors manually checked  each paper for adherence to the final inclusion criteria, resulting in 20 papers that intersect our desired research areas \textit{and} discuss developer guidelines for addressing accessibility issues. In addition to these 20 identified primary studies, we also examined Apple's and Google's design guidelines related to accessibility~\cite{AppleAccess,GoogleAccess}, as several of our primary studies referenced these sources.

After the search process concluded, one author extracted all accessibility guidelines discussed in the papers and platform documentation, and after the process, three authors met to discuss and verify that all guidelines were properly extracted in a joint meeting.  This process resulted in the derivation of 18 \textit{accessibility design guidelines}, which are illustrated in Table~\ref{tab:guidelines}. In this table, we provide (i) a short description of the guideline (with full definitions and examples available in our online appendix), (ii) the primary affected user groups, (iii) the sources that described the guideline, (iv) whether or not automated support for guideline has been implemented in past commercial or research developer tools, and (v) the guidelines targeted by \MotorEase. \textbf{It should be noted that none of the guidelines that M{\small OTOR}E{\small ASE} targets have been explicitly targeted by prior tools.} While touch-target size has been explored in prior work~\cite{Alshayban20, Abascal11, Kane11, GoogleAccess}, \textit{visual} touch-target size, which is of critical importance for motor-impaired users~\cite{Kong21}, has not been explored. While there is some recent work on Keyboard Accessibility failures in web applications that could affect Motor-impaired users~\cite{Chiou:CHI23}, this work does not explicitly target any of the guidelines targeted by \MotorEase. The Groundhog~\cite{Salehnamadi:ASE'22} tool is also inadvertently able to detect \textit{some} expanding section closures. The lack of exploration of these guidelines is largely due to the fact that implementing tools that detect when such guidelines are \textit{not} followed requires new types of automated UI screen understanding.

The four accessibility guidelines targeted by \MotorEase are (i) \textit{\textbf{Visual Touch Target Size}}, (ii) \textit{\textbf{Persistent Element Location}}, (iii) \textit{\textbf{Expanding Section Closure}}, and (iv) \textit{\textbf{Visual Icon Distance}}. One of these guidelines (\textit{Visual Touch Target Size}) was identified from a prior accessibility study, whereas the others come directly from Apple and Google's accessibility deign guidelines for mobile apps~\cite{AppleAccess,GoogleAccess}. These four guidelines were chosen as they had not been implemented by past work and could easily integrate with automated input generation (AIG) tools for Android, which is an important practical component of \MotorEase as we explain in Section~\ref{sec:approach}.  
In the following subsections, we describe each motor-impairment accessibility guideline in detail, and the screen understanding challenges in detecting guideline violations. More detailed descriptions of all  guidelines can be found in our online appendicies~\cite{appendix,site,zenodo}. 

\begin{figure}[t]
    \centering
    \includegraphics[width=0.45\textwidth]{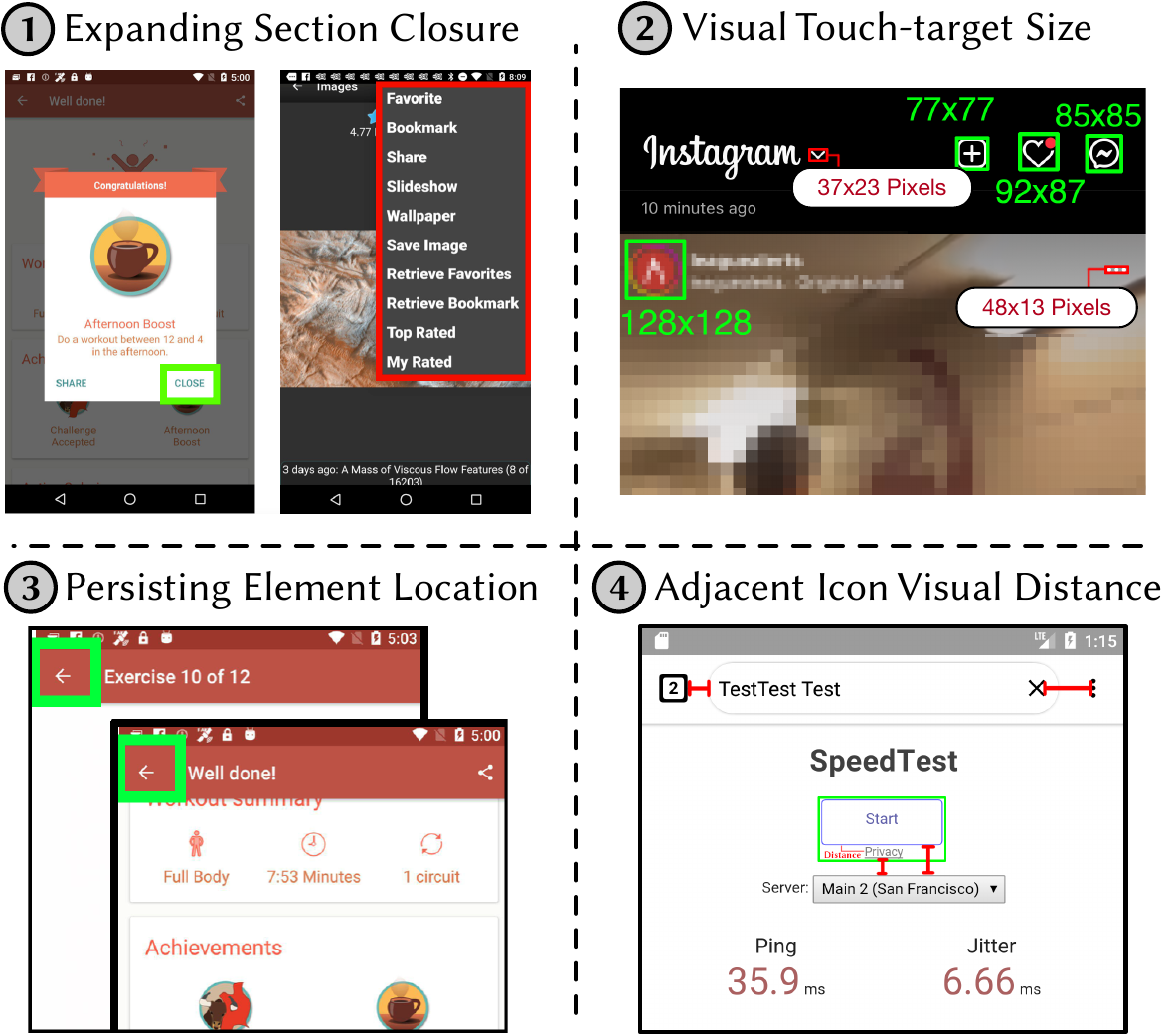}
    \caption{Illustration of four studied accessibility guidelines}
    \label{AllDetectors}
\end{figure}

\subsection{Expanding Section Closure}
Pop up menus and modal dialogs can provide meaningful information to the user, but closing them can be non-trivial for motor-impaired users who utilize a switch, as they may not contain explicit UI elements for closing the menu or dialog. As such, UI design guideline advocated for by both Apple and Google~\cite{AppleAccess,GoogleAccess} state that such closure UI elements should be present and easily interactive. Many expanding sections can be closed through a swiping gesture to dismiss the section or an external tap on the screen not within the bounds of the expanding section. Both of these options pose an accessibility issue for motor-impaired users due to the need for gestures and assumptive tapping on non-intuitive screen locations. A violation and adherence to this guideline is illustrated in Figure~\ref{AllDetectors}-\circled{1}. Detecting UIs that violate this design guideline can be difficult as it requires automatically identifying (i) whether a pop-up menu or modal dialog is present within a given screen, and (ii) whether or not a UI element supports closing the pop-up. 

\vspace{-0.5em}
\subsection{Visual Touch-Target Size}
Motor-impaired users who experience tremors in their hands can experience difficulty tapping precisely on icons. This makes it difficult for them to interact with elements as intended. Apple and Google suggest minimum UI element sizes of 44x44 pixels and 48x48 pixels respectively, such that users with minor motor impairments can more easily tap icons~\cite{AppleAccess,GoogleAccess}. 
Typically, the ``size'' of a UI element is defined by the \textit{touchable area} of that element, and not the \textit{visual area} occupied by the pixels of a given element. However, as stated above, it is important to provide sizable \textit{visual} touch targets to users with more limited motor control, such that they can better hone their more limited movements to tap desired UI elements. Most past work that aims to identify icons or UI elements that do not meet a given minimum threshold read UI metadata to examine the ``touchable area'' only, even if the visual size of the icon does not fill the entire area. Thus, this can create a disconnect between the \textit{visual} touch target size, and the \textit{touchable area}. An example of this is shown in Figure \ref{AllDetectors}-\circled{2}. The \includegraphics[width=0.04\linewidth]{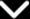} icon next to the Instagram logo has a touch target size of larger than 44x44, however, the visual size of this icon is quite small, a making it difficult to tap. 

\vspace{-0.5em}
\subsection{Persistent Element Location}

Applications link various screens together to aid users in completing complex tasks, however, certain UI elements need to exhibit \textit{consistent} placement to assist switch users with anticipating UI element scanning. This guideline specifies that icons that appear across multiple screens should appear in the same general area of the screen~\cite{AppleAccess,GoogleAccess}. This means the locations of elements such as back buttons or search icons that appear across multiple different screens should remain consistent. An example of a back button with a consistent location is illustrated in Figure~\ref{AllDetectors}-\circled{3}. Violations of this guideline can be difficult to detect as it requires automatic identification of corresponding UI elements across screens which may exhibit visual variability (e.g., displayed on different backgrounds).

\vspace{-0.5em}
\subsection{Adjacent Visual Icon Distance}

The design and placement of interactive icons that signal functional affordances is critical to ensuring a positive user experience, particularly for individuals with motor impairments. For users who may struggle with fine motor movement, it can be difficult to tap a single location on the screen without accidental triggers of other areas~\cite{Kong21}. As such, the \textit{Adjacent Visual Icon Distance} guideline states that adjacent ``clickable'' UI elements should be positioned at least eight pixels apart from one another.  This can be challenging as it again requires the automated inference of the visual area occupied by different UI elements. An example of this is shown in Figure~\ref{AllDetectors}-\circled{4}, wherein the ``Start'' and ``Privacy'' elements are located too close to one another and may result in accidental, unintentional triggering of the components by a motor-impaired user.

\section{The M{\large OTOR}E{\large ASE} Approach}
\label{sec:approach}

\begin{figure}[t]
	\centering
    \includegraphics[width=0.42\textwidth]{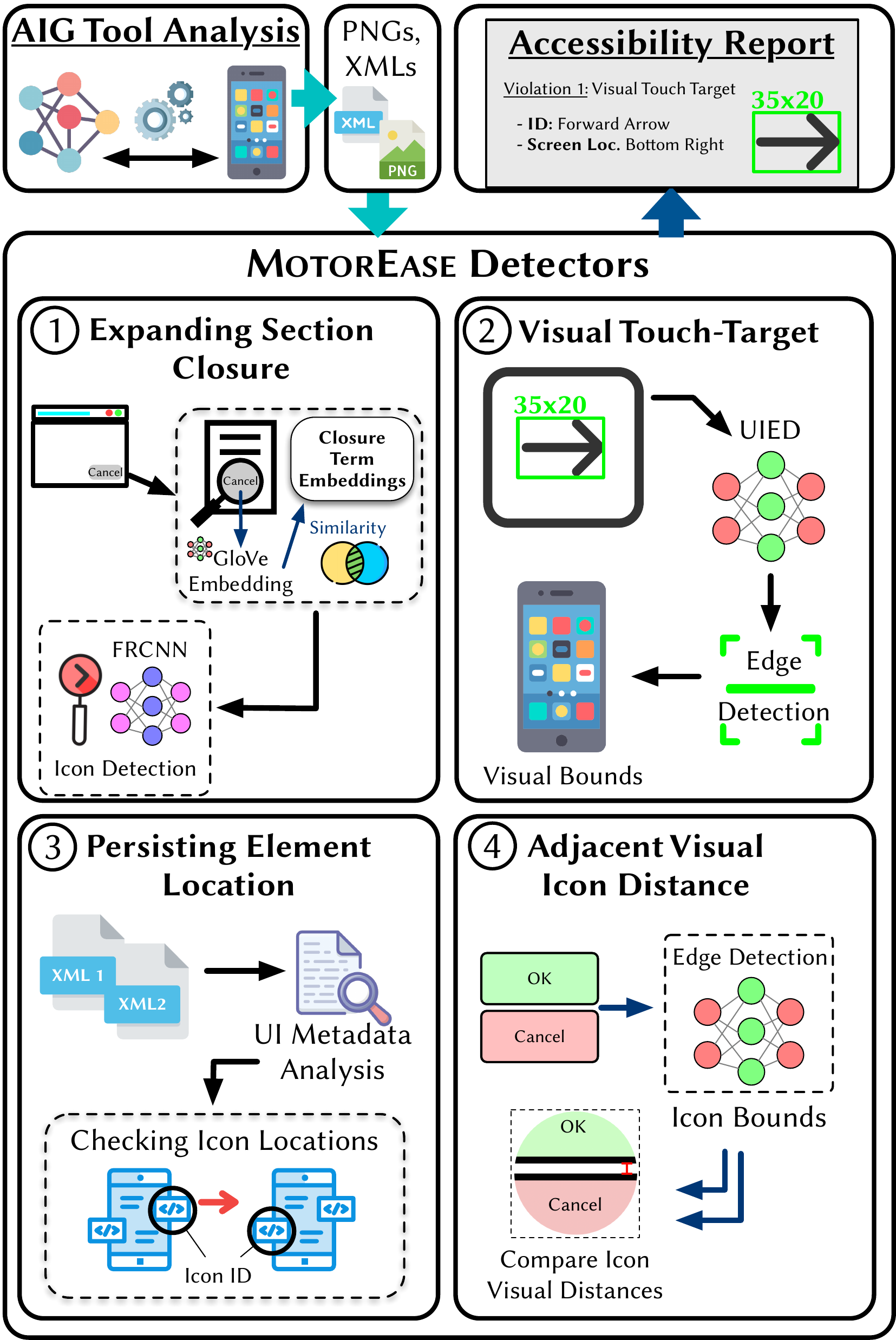}
	\vspace{-1em}
    \caption{Overview of \MotorEase's Workflow}
    \label{fig:overview}
\end{figure}

\MotorEase is an automated approach that aims to detect motor-impairment accessibility guideline violations by analyzing UI metadata and screenshots collected via automated input generation (AIG) tools (i.e., automated app crawlers, UI testing tools). MotorEase operates in three stages, and implements four guideline violation detectors, as depicted in Figure ~\ref{fig:overview}. First, an AIG tool is run on a target application to produce a set of screenshot and \texttt{\small uiautomator XML} files (\ie UI metadata) before and after each AIG tool action. We tailor our approach to utilize UI metadata generated using the \texttt{\small uiautomator} framework, which captures UI layout information in a structured \texttt{\small XML} format, as this is most prevalent utility used by recent Android AIG tools~\cite{mao2016sapienz,li2017droidbot,Gu:ICSE'19,Moran:ICST'16,crashscope,Su:FSE'17,Linares:MSR15,Linares:ICSME'17,Linares:ICSME'17-2,Zhao:FSE22}. It should be noted that \MotorEase does not require any pre-existing test cases, but instead can be used in conjunction with any of the AIG tools listed above. Second, \MotorEase utilizes a series of four \textit{violation detectors} to analyze the screenshots and UI metadata to determine if the target application failed to follow motor-impairment guidelines. Finally, \MotorEase collects the information from the detectors and compiles an \textit{accessibility report} that informs developers of accessibility guideline violations.

\subsection{Detectors}

The core components of \MotorEase are its four accessibility guideline violation detectors. Detectors \circld{1}, \circld{2} and \circld{4} operate upon \textit{single} \texttt{\small uiautomator} \texttt{\small XML} files, screenshots, in the form of \texttt{\small PNGs}, or both. Detector \circld{3} takes as input a series of \textit{multiple} \texttt{\small XML} and screenshot pairs. In the remainder of this section, we describe the technical underpinnings of each of \MotorEase's detectors.

\subsubsection{\textbf{Expanding Section Closure Detector}}

\begin{figure}[t]
    \centering
    \includegraphics[width=0.42\textwidth]{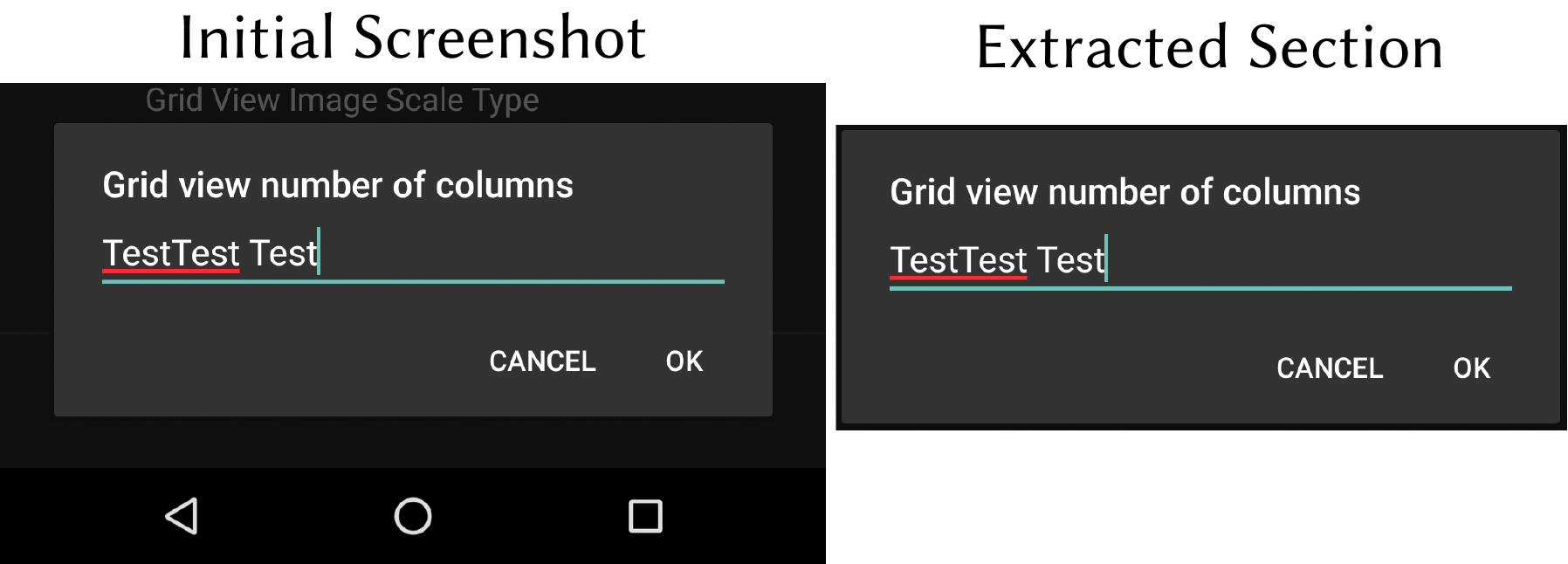}
    \caption{Extracted Expanding Sections}
    \label{extractedSec}
\end{figure}

The expanding sections detector aims to identify pop up messages or slide-in views that lack a visible means to close the section. Objects or text that imply closing the section is what \MotorEase aims to detect, if it cannot detect these, then a given screen with a dialog box or section is considered to be in violation of the guideline. 
This detector begins by determining whether the screen has an expanding section and then extracting it from the screenshot. This is done by identifying the largest element on the screen. \MotorEase extracts the largest \emph{android.widget.FrameLayout} and the largest \emph{android.widget.ListView} on the screen whose size is not the entire screen as the pop up screen or the slide-in list menu. An example of this is shown in Figure \ref{extractedSec}.

\begin{table}[t]
\centering
\small
\renewcommand{\arraystretch}{1.3}
\caption{Mapping of Sample Lexical Patterns to detect Closure}

\label{tab:text-patterns}
\begin{tabular}{p{8.1cm}}
\hline
\textbf{Initial Closure Words}\\

"close", "cancel", "dismiss", "done", "ok", "finish", "return"\\
\hline
\textbf{GLoVE Embedding Words}\\

 "deny", "allow", "exit","end","terminate","quit","back","stop","ignore", "proceed","save","apply","submit","confirm","abort","decline","reject","ignore"\\
\hline
\end{tabular}
\end{table}

Once \MotorEase has extracted each of the expanding sections, it then aims to determine whether the pop-up or section provides a clear means of closing it. In order to offer a robust solution, \MotorEase accomplishes this via two main procedures: (i) Semantic Text Matching and (ii) Icon Detection. This is due to the fact that closure controls can have either textual (\ie the word ``exit'') or visual (\ie an $\times$ icon) signifiers that indicate functionality.

\noindent\emph{\underline{Semantic Text Matching:}} \MotorEase's semantic matching technique defines a certain set of keywords that are likely to signify an element that can close an expanding section or pop-up. 
These keywords comprise common lexical patterns derived through two authors of the paper examining expanding sections that appear in 1500 randomly sampled screenshots from the RICO dataset~\cite{Deka:UIST'17}. The RICO dataset, comprising 9,000+ Android apps and 66,000+ screenshots, serves as a popular resource for mobile app research given its diverse set of screens and apps. We randomly sampled 1500 screens as it represents a statistically significant sample size of the ~66k screenshots present in the RICO dataset (95\% confidence level and 2.5\% margin of error). These words were terms that implied a closure or completion action, i.e. ``close'', ``dismiss'', ``cancel'', ``ok''. To ensure that the selected words for semantic matching were relevant, we employed a manual process in which two authors reviewed the randomly sampled dataset for expanding elements and words that implied closure. The resulting set of words was agreed upon by the authors as suitable for the task. To further expand this set of words, we further utilized GloVe embeddings~\cite{glove} to compare the original selected words to the entire dataset. Glove embeddings capture semantic relationships between words by considering global word co-occurrence patterns, resulting in dense vector representations that preserve meaningful similarities between words~\cite{glove}. We extracted additional words that exhibited a cosine similarity of at least 0.95 to the GloVe embeddings of the original selected words. This approach enabled us to carefully curate a comprehensive set of words for semantic matching that accurately represented closure. We provide the complete list of 25 closure words in Table~\ref{tab:text-patterns} -- as our experiments illustrate, we found these words to generalize well to our experimental benchmark.  

To extract the text from each image, \MotorEase leverages a combination of Google's Optical Character Recognition (OCR) text extraction~\cite{ocr}, which is based on the EAST OCR technique~\cite{zhou2017east} and the text present in the \texttt{\small uiautomator} \texttt{\small XML} file. We use both methods of text extraction as text displayed on the screen via images is often not captured in the \texttt{\small uiautomator} \texttt{\small XML} files. This method provides a binary classification for the presence of text that indicates a means to close the pop-up. If there are no matching extracted terms, \MotorEase then proceeds with icon detection.

\begin{figure}[t]
    \centering
    \includegraphics[width=0.42\textwidth]{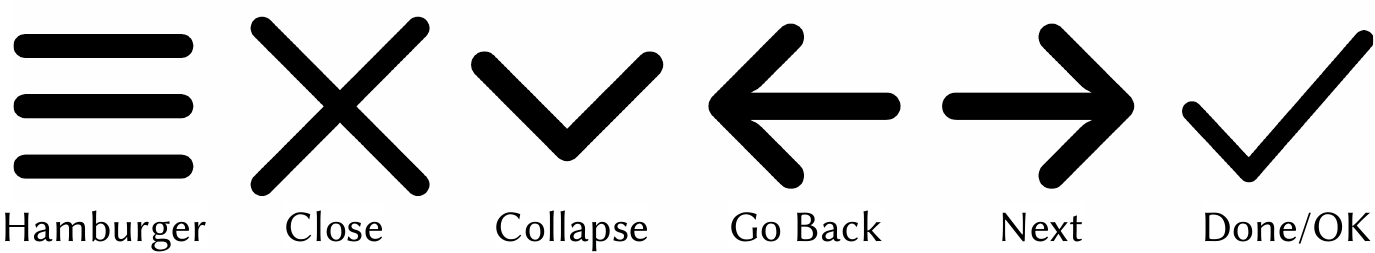}
	\vspace{-0.75em}
    \caption{Base Closure Icons}
    \label{fig:icons}
\end{figure}

\noindent\emph{\underline{Icon Detection:}} Icon detection is used to identify specific closing icons, i.e. $\times$, hamburger icon, checkmark, etc. Two authors examined the same set of 1500 screens from the RICO dataset discussed above and compiled a set of base icon types to train an image detection model in order to detect these icons. The chosen base icons are shown in Figure~\ref{fig:icons}. To accurately perform this detection, we trained a neural object detection model on a diverse set of examples of these identified icons, through a process we describe in detail below.

Training a neural object detection model typically requires a large-scale dataset with annotated examples of the target icons. To derive such a dataset, we \textit{automatically} constructed a realistic, synthetic dataset of icons that represent menu/pop-up closing. To do this, we extracted icon images with transparent backgrounds from the Fontawesome~\footnote{\url{https://fontawesome.com}} and Flaticon~\footnote{\url{https://www.flaticon.com}} image repositories until five icons for each icon type were identified that varied in color and style. Note that these icons do not directly appear in our evaluation dataset. We then superimposed these icons into random locations on screenshots derived from the RICO dataset~\cite{Deka:UIST'17}. During the analysis of the 1500 RICO screenshots two authors determined that a majority of icons on the screen were mathematically smaller than 10\% of the total screen size and larger than 2\% of the screen size. The screen size in question are the 1920x1080 pixel dimensions of the screen, limiting the maximum icon size to 192px and minimum icon size to 38px. Therefore, when an icon was superimposed on a screen, we varied their size between 2\% and 10\% of the screen area so that they remained relatively similar to the screens in the dataset. In this manner, we generated a dataset of 7,291 images (separate from the 1500 images sampled earlier) all with labeled and fully-localized icons on them (one per screen, spread evenly across the variations of our six icon types) and divided this into training and testing sets following an 80/20 split, 5,832 images for training, 1,458 for testing. We then used this dataset to train a Faster-RCNN object detection technique~\cite{ren2015faster} using the the torch-vision API~\cite{torchvision}. We generated $\approx$ 7k images as past work that uses a similar approach for icon detection was able to train an accurate model with this scale of data~\cite{Cardenas:ICSE'20,Cardenas:TSE23,Havranek:ICSE'21}. Our trained model was able to achieve over 95\% accuracy on the test portion of our dataset. 

\MotorEase passes the cropped out expanding sections to the model in order to detect the icons. 
This detector works by first checking for semantic text matches and if there is no match, it then checks for icons. \revision{\MotorEase performs text pattern matching first because of the potential for X icons on the expanding section \textit{not} related to closing the icon. Had we done the icon detection first, the X icons to delete text in the text-fields would have been detected by the object detector. This means that this screen would have been classified as closable even though the X icons do not imply closing of the section. Therefore we use both the text and images on the expanding sections to try and classify if it can be closed. }
If neither technique can pick up on a pattern or icon, \MotorEase reports the section as a violation to the expanding section guidelines, capturing the the screenshot name, FrameLayout/ListView name, and violation \revision{and make it known to the developer}.

\subsubsection{\textbf{Visual Touch-Target Detector}}

\begin{figure}[t]
    \centering

    \includegraphics[width=0.15\textwidth]{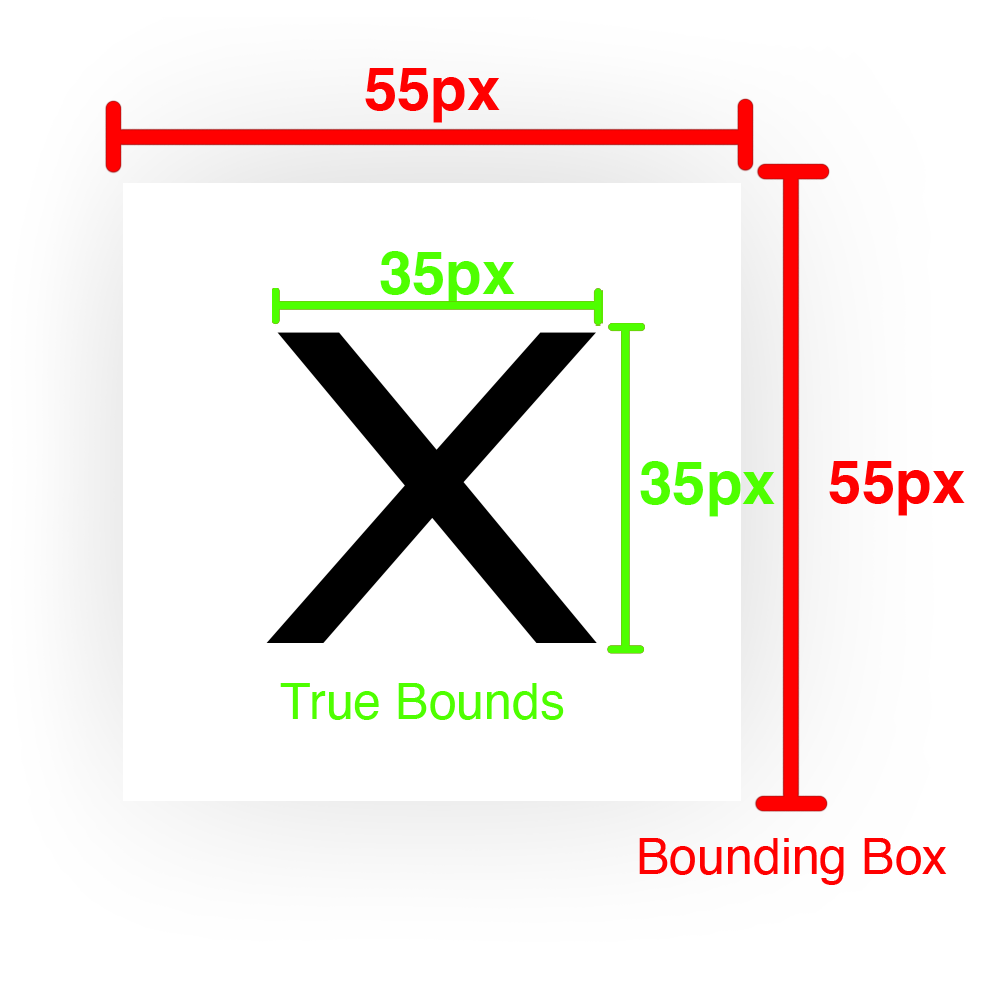}
\vspace{-1em}
    \caption{Example of Bounding Box vs. True Bounds}
    \label{boundBox}
\end{figure}

This detector uses both the screenshot and its corresponding XML file. It starts by processing the XML file and extracting the bounds for all elements which are clickable. \revision{XML has various properties in its metadata to describe an element, and if an element has a \emph{"True"} in the \emph{"clickable"} field, we determined that it is meant to be clicked on  or interacted with.} 
This detector aims to identify elements that have a visible area that is smaller than their tappable/clickable bounding box. Hence with this detector, \MotorEase aims identify elements whose \textit{visual} size are under 48x48 pixels~\cite{ANDRDesign}, even when the reported touchable area (as indicated by \texttt{\small uiautomator}) may be larger than 48x48. \revision{The bounding boxes in \texttt{\small uiautomator XML} files can show a bounding box whose size is larger than the actual size of the icon.} 
An example of this is shown in Figure~\ref{boundBox}. The true bounds of an icon corresponds to the visible area occupied by the icon. This can make these bounding boxes appear to be guideline abiding since they are generally larger than the true, visible bounds of the icon they hold. 
In order to identify these cases, \MotorEase adds 15 pixels to the width and height of the bounding boxes of each clickable item to extract the entire icon, before cropping the enlarged icon from the image.

After \MotorEase extracts the element, it is used as input to an edge detection algorithm, implemented in the UIED tool~\cite{UIED}, which is an approach that combines both neural object detection and unsupervised edge detection to effectively segment mobile app UI screens. This edge detection procedure is able to derive the \textit{visual} bounds of a given UI element, and would identify the 35x35px edges for the "X" as shown in Figure~\ref{boundBox}.
This procedure is applied to clickable icons extracted on each screen. Once the true edges are identified, \MotorEase is then able to then able to compare these bounds to the reported element size in the \texttt{\small uiautomator XML} file. If it is determined that the true element width or height is less than 48 pixels, \MotorEase labels that individual icon as a violation. 

\subsubsection{\textbf{Persisting Elements Detector}}
The persisting elements detector aims to identify an icon on a screen whose functionality remains the same, but location changes across multiple screens. An example of this is the navigation bar at the bottom of most applications. We expect that if the navigation bar is at the bottom of the screen on one screen, if a second screen has a navigation bar it should also be at the bottom of the new screen. 
This detector requires the use of both screenshots and \texttt{\small uiautoamtor XML} files from multiple screens, as \MotorEase directly analyzes properties of UI elements in the \texttt{\small uiautoamtor XML} files and compares the visual UI element similarity across multiple files.

This detector parses all of the \texttt{\small XML} files for a given application and records all of the UI element IDs. Then, it collects the location(s) across all \texttt{\small XML} files for each individual element ID. If there was more than 1 instance of a given ID across multiple screens, \MotorEase checks to determine whether the location bounds were the same as well as checking if the elements within those bounds are visually similar (95\% similar according to pixel-based mean squared error). If they are not, it deems it a violation of the persisting element guideline. This ensures that \MotorEase is checking every element in the application while checking to see if the icons with similar IDs have similar visual properties.
If there is an element that appears across the \texttt{\small XML} files more than once, \MotorEase examines them to see if the location is the same for the element. 
It relays this information back to the developer by providing the ID of the violating element in the generated Accessibility report. 

\subsubsection{\textbf{Visual Icon Distance Detector}}

The icon distance detector is designed to analyze icons on a screen and determine whether the visual distance between any two icons is less than 8 pixels, which combines findings from Kong \etal~\cite{Kong21} regarding visual icon size/spacing and recommendations from Google's accessibility Guidelines~\cite{GoogleAccess}. 
This detector analyzes a XML and screenshot pair, and extracts all of the \emph{clickable} elements on the screen using the \texttt{\small uiautomator} metadata. Once the clickable elements are identified, \MotorEase crops out the icons and sends them as input to UIED~\cite{UIED} to find the true visual bounds of the icon via edge detection. The true bounds are then used to subtract the padding from the initial touch target to the visual touch target. These modified bounding box values are then stored, and the detector iterates through all of the other icons on the screen to calculate the distance between the target UI elements and all other elements. The distance between the bounding boxes is calculated by determining the horizontal, vertical, or diagonal distance in pixels between two bounding boxes on a screen. Once the distance between the icons is computed, \MotorEase checks whether there is a distance is less than 8 pixels. If a violation is detected, \MotorEase includes a description of the UI elements in the generated accessibility report.  

\subsection{Accessibility Report Generation}

The accessibility report is generated using the target app screenshots that contain violations and a generated markdown file, with filenames to specific images of violations for each detector. We developed textual templates for each type of violation that are used by the report generation engine to automatically describe the accessibility violations by filling in the templates with information the \MotorEase analysis. This accessibility report aims to provide a comprehensive account of accessibility issues within a given app so that developers are able to take this information to make any changes they may need in order to make their apps more accessible.

\section{Design of Empirical Evaluation}
\label{sec:eval}

\newlist{questions}{enumerate}{2}
\setlist[questions,1]{label=RQ\arabic*.,ref=RQ\arabic*}
\setlist[questions,2]{label=(\alph*),ref=\thequestionsi(\alph*)}

In this section, we describe the procedure we used to evaluate \MotorEase. To achieve our study goals, we formulated the following five research questions: 
\begin{itemize}
		\item{\textbf{RQ$_1$} \textit{How accurate is the Expanding Section detector?}}
		\item{\textbf{RQ$_2$} \textit{How accurate is the Visual Touch Target detector?}}
        \item{\textbf{RQ$_3$} \textit{How accurate is the Persisting Element detector?}}
        \item{\textbf{RQ$_4$} \textit{How accurate is the UI Element Distance detector?}}
		\item{\textbf{RQ$_5$} \textit{Does \MotorEase identify a limited number of false positive and negative violations?}}
\end{itemize}
        
\subsection{RQ$_1$ - RQ$_4$: Violation Detection Capability}
 
\MotorEase is one of the first tools to support the detection of violations of accessibility design guidelines targeting motor-impaired users, and accomplishes this via understanding the visual and textual modalities Android UI screens. Given that prior techniques are not able to explicitly detect the design violations that \MotorEase targets, by using the visual comprehension of the screen, we both derived an entirely novel benchmark and designed an evaluation methodology that tests each of its four detectors individually to determine their accuracy, precision, and recall. In addition, we also examine the false positive and false negative rates to better understand the practical utility of \MotorEase. The evaluation metrics used in this study provide insights into the ability of each detector to identify true positives and true negatives. Accuracy gives us an overall ability to deduce each detectors ability to detect true positive (TP) and true negative (TN) values accurately. Note that we balance the positive and negative samples in the \MotorCheck benchmark to allow for an informative accuracy measurement.
	
	\begin{center}$Accuracy = \frac{TP+TN}{TP+TN+FP+FN}$\end{center}
	
\noindent In the context of our study, a True Positive ($TP$) is defined as the detection of an existing design guideline violation as defined in our ground-truth dataset. A false positive ($FP$) is defined as the detection of a violation when one does not exist. A False Negative ($FN$) occurs when our approach does not report a violation, but one exsists in the ground truth. Finally, a True Negative ($TN$) occurs when the approach does not report a violation and one does not exist. In addition to accuracy, we also measure the precision and recall (as defined below) to provide a more complete picture of \MotorEase's performance. The results of \MotorEase were manually validated (\ie two authors compared MotorEase's output to the ground truth.)
	
	\begin{center}$Precision = \frac{TP}{TP+FP}$ \hspace{1em}$Recall = \frac{TP}{TP+FN}$\end{center}

\revision{\noindent Given these two evaluation metrics, we can determine how accurately each detector works. \MotorEase's overall effectiveness can be derived by averaging the the accuracy/precision/recall across all four detectors, providing a comprehensive understanding of \MotorEase's ability to detect accessibility guideline violations. }

\subsection{RQ$_5$: M{\normalsize OTOR}E{\normalsize ASE}'s Practical Utility}

In order to investigate \MotorEase's practical utility, we also report both the false positive and false negative rate, as these reflect the need for a developer to sift through incorrect violation reports, or lost quality in terms of miss unreported violations. This helps to provide a more holistic picture of \MotorEase's performance.

	\begin{center}$False Positive Rate = \frac{FP}{FP+TN}$ \hspace{1em}$False Negative Rate = \frac{TN}{FP+TN}$\end{center}

\begin{table}[t]
	\small
	\begin{center}
	\caption{Expending Sections Detector Test Dataset}
	\begin{tabular}{ c|c|c|c } 
		\hline
		\textbf{Total Files} & \textbf{Close: Icon} & \textbf{Close: Text} & \textbf{Cannot Close} \\
		\hline
		483 & 27 & 214 & 242\\ 
		\hline
	\end{tabular}
	\label{t2}
	\end{center}
\end{table}

\begin{table}[t]
	\begin{center}
	\small
\vspace{-2em}
	\caption{Visual Touch-target, Persisting Elements, and Visual Icon Distance Test Datasets}
\vspace{-1em}
	\begin{tabular}{ c|p{.6in}|p{.53in}|p{.5in} } 
		\hline
		\textbf{Detector} & \textbf{Total Files/Apps} & \textbf{Violations} & \textbf{Non-Violations} \\
		\hline
		Touch-Target & 400 files & 176 & 224 \\
		\hline
		Persisting Elements & 49 apps & 24 & 25 \\
		\hline
		Icon Distance & 400 files & 42 & 358 \\
		\hline
		
	\end{tabular}
	\label{t1}
	\end{center}
\end{table}

\subsection{Derivation of the M{\normalsize OTOR}C{\normalsize HECK} Benchmark}
\label{subsec:dataset}

Given that no prior approach has targeted the motor-impairment design violations targeted by \MotorEase, we develop a novel benchmark called \MotorCheck which we discuss below. It should be noted that all of the accessibility violations of this benchmark are real, no synthetic violations were injected in it's construction -- instead real violations were rigorously manually annotated.

To derive the initial set of screenshots and xml files for \MotorCheck we applied the {\sc CrashScope}~\cite{crashscope} automated testing tool to 70 popular Android apps that are cross-listed on both FDroid and Google Play. To do this, we gathered a list of apps from F-Droid~\cite{Fdroid} and considered only those apps that were cross-listed on Google Play~\cite{GooglePlayStore} and had at least 1000 downloads -- providing some confidence in the popularity of the chosen applications. We provide a full list of these applications with download statistics and links in our online appendicies~\cite{appendix,site,zenodo}. During this process, we used one of {\sc CrashScope}'s exploration strategies to extract 2,864 screenshot/\texttt{\small XML} pairs. Note that the goal of our study in assessing \MotorEase's capabilities is independent of the coverage provided by the underlying testing tool, although {\sc CrashScope} has been illustrated to be competitive with other tools~\cite{Moran:ICST'16}. It should be noted that the screen coverage of \MotorEase is dependent upon the Android AIG tool that the approach is paired with. Given recent advances in AIG tools~\cite{wang_vet_2021}, \MotorEase can integrate with these new tools and take advantage of the improved coverage. Next, we describe the dataset derivation process for each detector. 
Note, given that the data labeling process is quite objective for violations of our identified guidelines, for each dataset, we had one author manually label each instance, and another author verified the results. During this process, no instances of conflicts were noted, again due to the largely objective nature of the labeling procedure. We provide an overview of the \MotorCheck benchmark data in Tables~\ref{t2}~\&~\ref{t1}.
 
\subsubsection{Expanding Section Closure Detector Dataset} In order to detect expanding sections, this detector requires an input screenshot and its corresponding \texttt{\small XML} file. In order to evaluate the detector and remove any bias, one author labeled expanding sections without closure elements until all CrashScope files were exhausted, resulting in 241 screens. Of the 241 screens that had an expanding section, there were 121 screens were FrameLayouts and the remaining 120 were ListViews. Then, in order to balance the dataset, an additional 242 screenshots without violations were randomly selected to complete the dataset, for a total of 483 screens. The additional screens without violations consisted of expanding sections that could be closed. Table~\ref{t2} provides more information on how the dataset was split between violations and non-violation samples. Labeling was done manually using LabelStudio~\cite{LabelStudio}. For the screenshots without violations, icon types and closure word types were labeled. If neither the icon nor text clearly showed a means of closing the section, it was labeled as a violation. 

\begin{table*}[t]
\vspace{-1em}
\footnotesize

    \caption{Overall Results for \MotorEase Detectors \& Baselines}
    \centering
    \begin{tabular}{ c|c|c|c|c } 
        
         \textbf{Approach} & \textbf{Precision} & \textbf{Recall} & \textbf{Accuracy} & \textbf{F1-Score}\\\hline
        \textbf{{\sc\textbf{MotorEase}} (Viz T.-Target)} & 1.0000 & 0.6648 & 0.8525 & 0.7986\\ 
		\textbf{G-Accessibility Scanner (T.-Target)} & 0.5556 & 0.5085 & 0.6025 & 0.5310 \\\hline 
        \textbf{{\sc\textbf{MotorEase}} (Exp. Sec)} & 0.9042 & 0.9205 & 0.9123 & 0.9129\\ 
		\textbf{Groundhog (Exp. Sec)} & 0.6849 & 0.8659 & 0.7207 & 0.7648 \\ \hline
        \textbf{{\sc\textbf{MotorEase}} (Pers. Elem)} & 0.8214 & 0.9583 & 0.8776 & 0.8846\\
        \textbf{{\sc\textbf{MotorEase}} (Icon Dist)} & 0.7119 & 1.0000 & 0.9575 & 0.8317\\\hline
        \textbf{{\sc\textbf{MotorEase}} (All Detectors)} & \textbf{0.8594} & \textbf{0.8859} & \textbf{0.8999} & \textbf{0.8570}\\
       
    \end{tabular}
    \label{results}
\end{table*}

\subsubsection{Visual Touch-Target and UI Element Distance Detector Dataset} The touch-target detector requires screenshot and XML pairs to determine if the screens had a touch-target violation (\ie the \texttt{\small XML}) and visual bounds differed. In order to evaluate the detector and remove any biases, we randomly chose 400 screenshots and XML pairs generated by {\sc CrashScope} stratified across our 70 applications. This sample size was used as it represents a statistically significant sample of the total number of extracted {\sc CrashScope} screens (99\% confidence interval). Table~\ref{t1} provides the dataset splits between violations and non-violation samples. Labeling for these screenshots was performed manually. One author analyzed each of the screenshots and set a bounding box on each of the interactive elements on the screen using Label Studio~\cite{LabelStudio}. If the size of the bounding box was less than 48 in width or height, it was labeled it as a violation, else it was labeled as a screen without violations. Additionally, the author checked the distance between all components on these screens and labeled any instances where UI elements were less than 8 pixels apart. This was done for all 400 images.

\subsubsection{Persisting Elements Detector}

The persisting elements detector requires multiple app XMLs in order to detect violations. The CrashScope dataset \cite{crashscope} contains 70 applications and their screenshots. We filtered out apps that only contained screenshots of similar screens, resulting in 49 applications, 24 of which had persisting elements that violated our guideline, and 25 of which adhered to our guideline. One author labeled each app as having a violation or not having a violation. During the labeling, this author also specified which specific screenshot exhibited the violation.

\subsection{Comparison to Baseline Techniques}

While the Accessibility issues that \MotorEase targets have not been explicitly targeted by past tools, there are two tools which are capable of detecting a subset of the accessibility violations identified by MotorEase. These two baselines are  Groundhog~\cite{Salehnamadi:ASE'22} and Google Accessibility Scanner~\cite{GoogleScanner}. We ran these two tools on the same \MotorCheck benchmark used to evaluate \MotorEase to keep the comparison fair and consistent, and we report the same metrics for both \MotorEase and the baseline techniques. Upon careful analysis of these baselines, Google Accessibility Scanner and Groundhog are only capable of detecting Touch-target size violations and Expanding Sections violations, respectively.

The first baseline tool we compared \MotorEase to in our updated evaluation is Google's Accessibility Scanner~\cite{GoogleScanner}. This tool operates directly upon the dynamic representation of the GUI as reported by \texttt{\small uiautomator}, and checks to whether the bounds of screen elements fall below the 48x48 dp threshold. To apply this tool, we launched the each app included in the MotorEase dataset on an emulator of the same screen dimension (1920x1080) and Android version as the screens in the MotorCheck benchmark  and manually navigated to the screen in question, and triggered the Accessibility Scanner tool. One author then manually compared the output results from Accessibility Scanner to the ground-truth visual touch-target size violations defined in \MotorCheck. We then reported the Precision/Recall/Accuracy and F1 Score Metrics.

The second baseline tool that we compared \MotorEase to is the Groundhog Accessibility tool~\cite{Salehnamadi:ASE'22}. We worked directly with the authors of the Groundhog tool, who were quite helpful after some initial issues initializing the tool, in order to apply Groundhog to the \MotorCheck benchmark. Groundhog functions by first sending touch-based actions to a given Android app screen running on an emulator or real device, and then attempts to exercise the same actions using one of Google's accessibility services, such as Talkback. Given the manner in which Groundhog works, the only Accessibility issue in \MotorCheck that it was applicable to is the Expanding Section guideline violations. To apply Groundhog to detect Expanding Section violations in \MotorCheck we launched a target on an Android emulator configured to the 1920x1080 screen size and Android version of the MotorCheck screens, and then ran Groundhog on the screen with an expanding section. Groundhog then navigated the screen both with and without assistive services to determine whether it could close the expanding section, if there was a discrepancy, this was reported by Groundhog. Then one author manually checked the output of Groundhog to determine if it was able to detect an expanding section violated the \MotorCheck guideline and could not be closed by a tappable element. We then reported the metrics seen in Table~\ref{results}

\section{Empirical Results}
\label{sec:results}

\begin{figure}[t]
    \centering
    \includegraphics[width=0.35\textwidth]{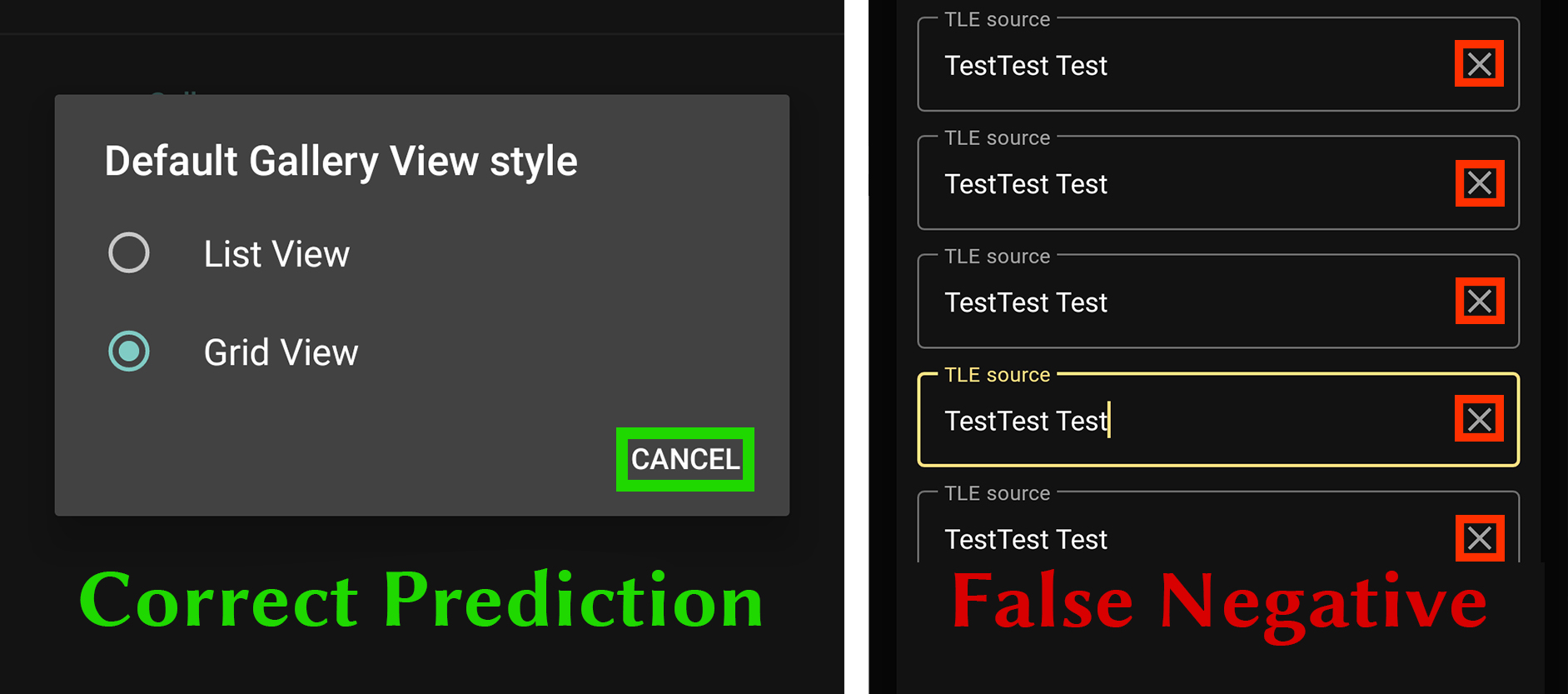}
	\vspace{-1em}
    \caption{Expanding Section Closure Detection}
    \label{expandingResults}
	\vspace{0.5em}
\end{figure}

\subsection{RQ$_1$: Expanding Section Detector Accuracy?}

The expanding section detector performed well across nearly all of our studied metrics, as indicated in Table~\ref{results}. With a precision of .9042, F1-Score of 0.9129, and an accuracy of 0.9123, this detector shows promising results that it is capable of identifying a section's ``collapsibility'' accurately. This indicates that, by using \MotorEase, developers will have an increased chance of identifying sections that were designed without a method of closure which may impede motor-impaired users. Importantly, as per Table~\ref{results} our approach surpassed the performance of Groundhog's ability to detect closable sections, with a precision of 0.6849, demonstrating its effectiveness in comparison. The difference between the two is attributable to the fact that \MotorEase and Groundhog do not have the same objective. MotorEase aims to detect the presence or absence of closing icons, while Groundhog aims to determine whether a closing icon can be accessed using an assistive service. If a closing icon is missing, Groundhog cannot detect it. Since there is no closing icon, Groundhog does not even attempt to close it using an assistive service. \revision{Groundhog relies on an accessibility service to detect icons on the screen, However, if the icon does not have any metadata indicating that it is an interactive, it is unable to interact with the icon and close it. \MotorEase's novelty lies in its ability to consider the visual presence of the icon independent of its metadata description. } However, \MotorEase's detector does struggle to detect certain instances in the \MotorEase benchmark. An example of a successful and unsuccessful prediction is shown in Figure \ref{expandingResults}. The example on the left side is an expanding section that the detector correctly identified as collapsible. It is correctly identified by \MotorEase as a closing section because of the semantic matching's ability to generalize the word "cancel" as a means of closing the section. The image on the right is not collapsible but the detector deduced that it was collapsible. This was due to the use of X icons in the training data for the object detection model. The "X" icons used in this image imply the deletion of text, but this detectors object detection model is also trained to identify "X" icons that may be used to close the expanding section. It should be noted that this was an outlier in our dataset, and that the pattern for detecting "X" icons generally worked as expected.

\subsection{\hspace{-0.8em}RQ$_2$: Visual Touch Target Detector Accuracy?}

The touch-target detector performed well as illustrated in Table \ref{results}. The detector exhibited perfect precision, an F1-Score of 0.7986, and an accuracy of 0.8525, showing encouraging results of its ability to detect and classify screens with touch target violations well.  Given the results, this detector is successfully able to give developers an insight into smaller icons that may inconvenience users with tremors and inaccurate touches. Importantly, our approach surpassed Google Accessibility Scanners's ability to detect visually small elements on the screen, which only had a precision of 0.5085, demonstrating its effectiveness in comparison. The primary reason for the gap in performance is that Google Accessibility Scanner is not able to check the \textit{visual} touch-target size, and can only parse the reported size from the \texttt{\small uiautomator} framework, which may not necessarily correspond to the visual touch target size. However, while nearly all violations returned by this detector are correct, it does tend miss certain types of violations. For instance, it cannot detect icons on the screen which are not labeled as clickable in the XML. By default, all elements on an android device have a clickable property with a boolean True/False label. If that property is labeled as False, \MotorEase does not consider the object to be clickable. Dynamically generated screenshots may not always have the metadata information for each element on the screen, and this absence of information was the main reason for incorrect or missed violations. This detector could be augmented in the future with work from the HCI community aimed at assessing icon tap-ability~\cite{Swearngin:CHI'19}.

\begin{figure}[t]
    \centering
    \includegraphics[width=0.30\textwidth]{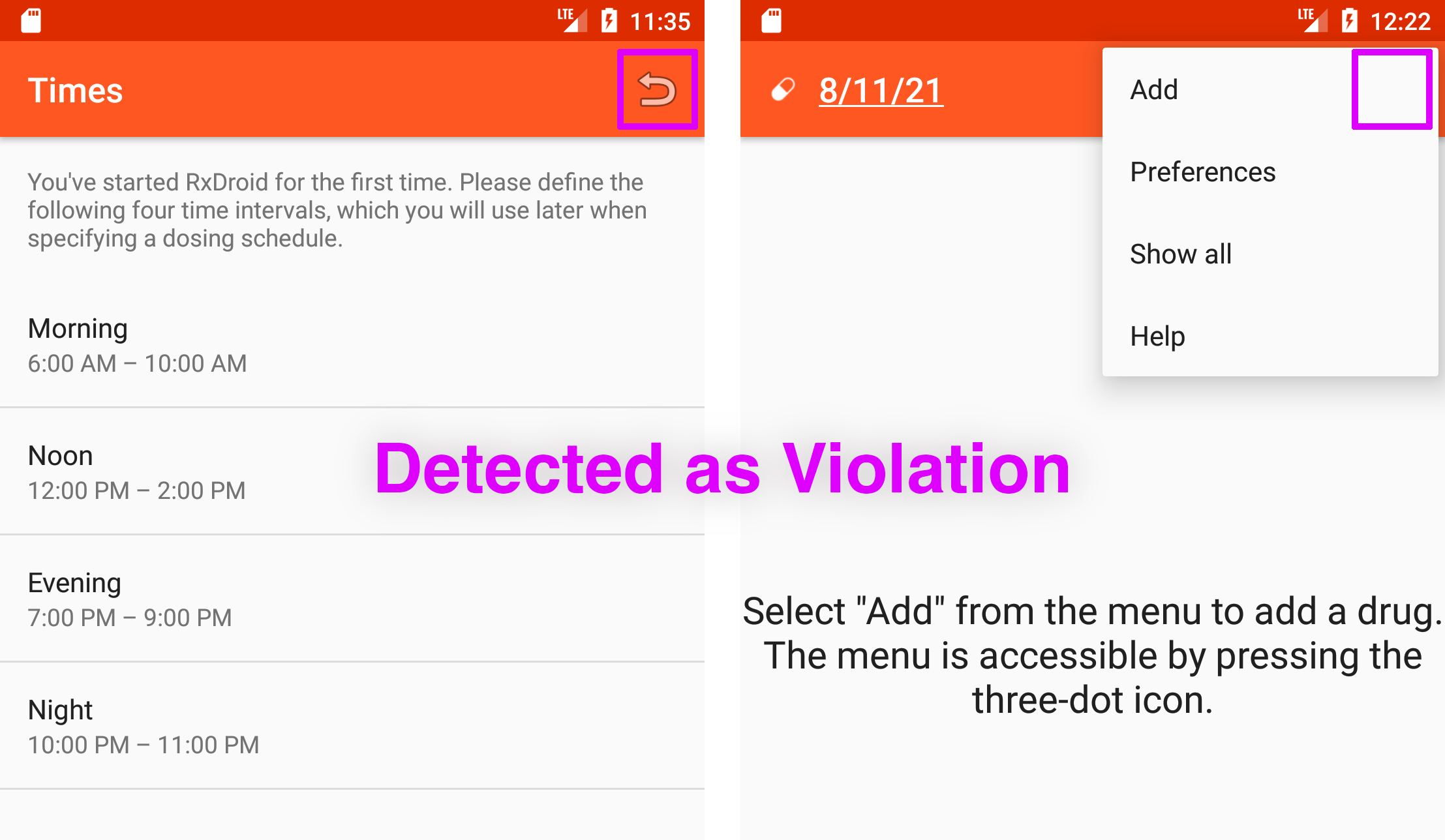}
\vspace{-1em}
    \caption{Persisting Elements Detection}
    \label{persistingResults}
	\vspace{1em}
\end{figure}

\subsection{RQ$_3$: Persisting Element Detector Accuracy?}

This results of this detector are presented in Table \ref{results}. With a precision of 0.8214, F1-Score of 0.8846, and accuracy of 0.8786, this detector shows encouraging signs of viability. This detector's recall rate of 0.9583 suggests that the detector identifies true positives at a high rate. This detector, however, relies heavily on the XML to locate elements on the screen, which can lead to mis-classifications. One such example is shown in Figure \ref{persistingResults}. The example shows the undo icon on the screenshot on the left and a menu on the right side where the undo icon would be. The XML for this second screen has the undo icon in the data, but its bounds and information are missing since they are not visible on the screen. This was classified as a violation though it is not a violation.

\subsection{\hspace{-0.8em}RQ$_4$: Visual Icon Distance Detector Accuracy?}
\begin{figure}[t]
    \centering
    \includegraphics[width=0.35\textwidth]{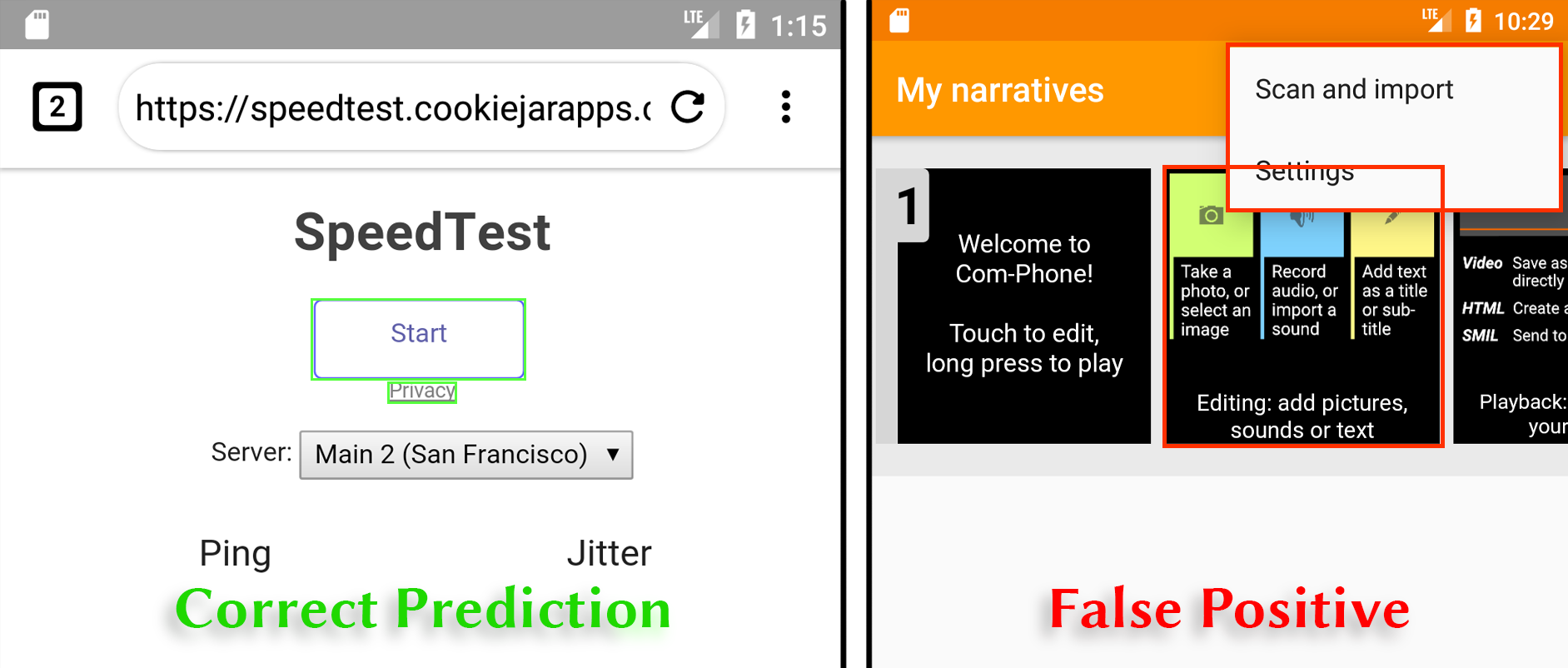}
    \caption{Visual Icon Distance Detection}
    \label{icondistanceResults}
	\vspace{1em}
\end{figure}

This results of this detector are presented in Table \ref{results}. With a precision of 0.7119, F1-Score of 0.8317, and accuracy of 0.9575. This detector achieved a perfect recall rate, suggesting that this detector provides developers with a reliable tool that is capable of accurately detecting closely placed icons, prompting potential UI design and icon placement adjustments. Like the Visual Touch-Target Violation detector, this detector relies heavily on the \texttt{\small uiautomator} metadata specifying clickable components, and the accuracy of the UIED element bound detector. The latter led to certain cases of inaccurate reporting of violations, due to incorrect overlapping bounds. \revision{Both of these examples are shown in Figure~\ref{icondistanceResults}. The first example shows a correct prediction where \MotorEase correctly identifies two icons on the screen and determines that they are not a minimum of 8 pixels in distance. However, Figure \ref{icondistanceResults} also has an example of a false positive prediction which shows an overlap of two elements on the screen. Due to the visual bounds of each overlapping, the distance between the two is 0, therefore predicting a false positive. }

\subsection{RQ$_5$: False Positives and Negatives}
\begin{figure}[t]
    \centering
    \includegraphics[width=0.50\textwidth]{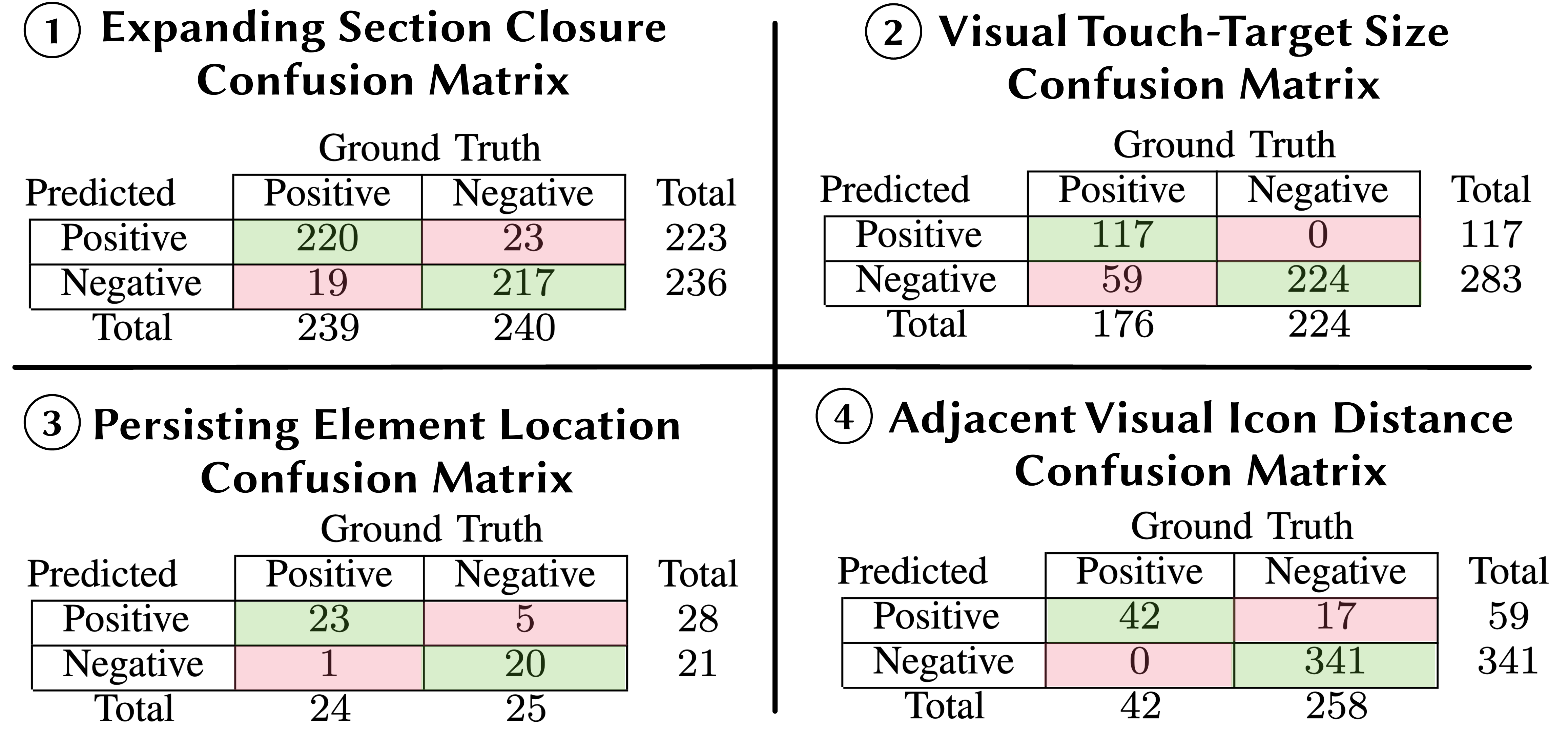}
	\vspace{-1em}
    \caption{Detector Confusion Matrices}
    \label{matrices}
\end{figure}

The confusion matrices for the detectors are shown in Figure~\ref{matrices}, where green boxes illustrate predictions that matched the ground truth, and red boxes illustrate predictions that did not match the ground-truth. \revision{These figures provide a visual representation of false positive and negative rates.} 
Figure~\ref{matrices}-\circld{2} illustrates that the visual touch target detector never produced a false-positive outcome. This is due to the fact that the detector specifically extracts elements labeled as clickable in the \texttt{\small XML}, therefore once they are extracted and have their edges analyzed, the detector is able to detect violations with certainty. The confusion matrix for the expanding section detector is shown in Figure~\ref{matrices}-\circld{1}, there were 23 false positive predictions, mainly due to limitations related to lexical pattern matching. Finally, \MotorEase's persisting element detector identified only 8 false positives, mainly due to inconsistencies in matching elements across screens due to unexpected changes in \texttt{\small uiautomator XML} files. In evaluating the effectiveness of \MotorEase, we also considered the impact of false negative predictions, as they signal violations that are not flagged by MotorEase, and hence could reach end-users. \revision{However, our evaluation revealed that the recall rate of \MotorEase was relatively high, indicating that it is a dependable tool with a high true positive rate, thus demonstrating its practical applicability.}
In regards to \MotorEase's low false negative prediction rate, The confusion matrix analysis for the icon distance detector shown in Figure~\ref{matrices}-\circld{4} shows that \MotorEase predicted 0 false negatives. Moreover, the confusion matrix analysis for the persisting elements detector shown in Figure~\ref{matrices}-\circld{3} showed only one false negative prediction. Similarly, the confusion matrix for the expanding section detector presented in Figure~\ref{matrices}-\circld{1} demonstrated a false negative rate similar to its false positive rate, further supporting the viability of MotorEase. Overall, our evaluation demonstrates that \MotorEase is likely a generally practical tool, exhibiting a relatively low rate of false positives and negatives. \revision{\MotorEase leverages it's ability to visually comprehend the visual and textual contents of a screen to determine accessibility violations. This approach offers distinct advantages. For instance, \MotorEase showcases its adaptability by effectively detecting accessibility guideline violations regardless of variations in UI design or format. In contrast, traditional metadata-based approaches might be limited in their ability to analyze physical elements on the screen.}

\section{Related Work}
\label{sec:rel-work}

\subsection{Accessibility Studies on Mobile Apps}

There exists a large body of work that aims to understand how users with disabilities use their devices, and the potential accessibility issues that exist in current software applications~\cite{Mchugh20,Almeida10,Yan19,Silva20,Santiago22,Mateus20,Aizpurua14,Silva19,Oh13,Montague14}. Such studies tend to take two forms, user studies~\cite{Aizpurua14,Mateus20} and empirical analyses of software~\cite{Silva20,Yan19}.

A study by Alshayban et al. \cite{Alshayban20} examined the current state of accessibility issues in Android applications~\cite{Alshayban20} using a Google-provided accessibility testing framework~\cite{GoogleAccess}. They found that Text Contrast, Touch Target, Image Contrast, and Speakable Text are the most frequent accessibility issues~\cite{Alshayban20}. Vendome et al. \cite{Vendome19} also examined the prevalence of accessibility issues in Android apps. In this study, the authors mined thousands of android applications and analyzed the usage of accessibility APIs and whether or not applications adhered to accepted guidelines. In addition, they mined thousands of messages on Stack Overflow and other interaction platforms to understand the sentiment of developers and the types of questions they were asking. They found that most accessibility based conversations were centered around LV features while a lesser number focused on DHH features. 

Our approach is motivated by and complements what researchers have discovered in the above studies. These studies have shown the importance of the problem that \MotorEase tackles through illustrating the widespread prevalence of accessibility issues in mobile apps and illustrating a comparative lack of awareness of motor-impairment design guidelines and the needs of such users. Hence, these studies both motivate and validate our work on \MotorEase.

\vspace{-1em}
\subsection{Accessibility Testing}

Software testing for accessibility aids developers in identifying violations of guidelines set forth by companies and governments~\cite{Norman13, GoogleAccess, AppleAccess, Park14}. A wide range of research has been carried out to automate this process~\cite{Ramachandra18,Brajnik15,Salehnamadi21,axeray,Norman13,Eler18,Salehnamadi:ASE'22}. We discuss the most closely related approaches below.

Eler et.al. introduced the MATE tool~\cite{Eler18} that uses automated dynamic analysis to check for accessibility issues that affect users with visual impairments in mobile apps, and generate detailed reports that facilitate developers fixing identified issues. Similar to MATE, \MotorEase also leverages automated dynamic analysis and is able to generate reports that aid developers in fixing accessibility issues. However, our tool is differentiated by the ML-based analyses employed, and by it's focus on motor-impaired users.

Latte~\cite{Salehnamadi21} is an accessibility testing framework introduced by Salehnamadi et.al. for android applications that aims to provide a deeper analysis compared to testing frameworks provided by Google~\cite{ANDRDesign, GoogleAccess} by testing how accessibility services, such as VoiceOver, function in conjunction with feature-based use cases.
The authors carried out an evaluation of their tool using the \texttt{\small switchAccess} and \texttt{\small TalkBack} services~\cite{AppleAccess, GoogleAccess} and found that several applications did not accommodate for both forms of accessible interactions. Latte is one of the only tools or studies that explicitly considers accessibility issues for users with motor impairments, given that it is capable of integrating with the \texttt{\small switchAccess} service in Android. However, given Latte's use case driven nature it both (1) requires pre-existing test cases, which many mobile apps have been shown to lack~\cite{Lin:ASE'20}, and (ii) cannot detect violations of the specific accessibility guidelines targeted by \MotorEase, as it attempts to carry out actions of a use case using an assistive service, and does not analyze the UI for specific patterns.  In short, \MotorEase and Latte serve largely \textit{complementary} purposes, that is \MotorEase provides UI design guidance to developers to avoid common pitfalls related to motor-impaired accessibility issues, and Latte can point out issues specific to given use cases and accessibility services. 

Chen et al. \cite{Chen22} introduced Xbot, an accessibility testing tool that is capable of identifying accessibility issues within an app using a combination of dynamic and static program analysis. Xbot is not able to uncover any of the accessibility issues targeted by \MotorEase. \MotorEase exhibits novelty as compared to Xbot as it  utilizes semantic understanding of the visual and textual elements of UI screens to detect new issues that affect motor-impaired users. 

Finally, recently Salehnamadi et.al introduced the Groundhog approach~\cite{Salehnamadi:ASE'22}, which is an accessibility crawler for mobile apps. Groundhog implements an automated UI crawler that explores an app both with and without assistive services, such as \texttt{\small TalkBack}, and notes any cases where an action can be performed through traditional touches, but cannot be performed via an assistive service. Again, \MotorEase is largely \textit{complementary} to this work, in that Groundhog targets general issues related to actionability and locatability of UI elements more broadly, but does not target the specific motor-impairment accessibility issues addressed by \MotorEase. In fact, \MotorEase aims to address two specific classes of issues identified in the Groundhog paper as being important for future work, (i) counterintuitive navigation (e.g., persisting elements) and (ii) inoperative actions (e.g., expanding sections).

\subsection{Accessibility-Based UI Comprehension}

Given that users interact with software through a UI, and past work has illustrated accessibility issues present in UIs~\cite{Salehnamadi21,Li22,Bajammal21,Peng19}, there is a body of work dedicated to automatically comprehending and augmenting UIs to identify and circumvent accessibility and UI issues~\cite{Liu21,Liu2021,Chen18,Gajos07,Shiver15,Montague12,Zhang21,Moran:ICSE'18,Moran:ASE'18}. 

UI elements and icons typically need to be labeled in order for screen readers to be able to properly describe their appearance and functionality, however, such metadata is often missing from apps~\cite{chen2020unblind}. Zhang et.al.\cite{Zhang21} and Chen et.al.~\cite{chen2020unblind} designed machine learning models trained from both existing UI labels and annotated labels from developers.
Follow-up work by Mehralian et. al. introduced COALA~\cite{Mehralian:FSE'21}, which aimed to improve upon the automated icon labeling by considering context related to screen text and region to build a multi-modal model for predicting icon labels.

Mansur et.al. introduced the \AidUI~\cite{mansur2023aidui} tool, which uses semantic screen understanding to automatically identify and localize Dark Patterns in mobile app and web user interfaces. This technique uses similar techniques for semantic screen understanding as \MotorEase, but in different ways. For instance, while AidUI uses element size and distance between elements as a factor in determining certain dark patterns, \MotorEase aims to compare to the programmatic element size and visual element size to detect accessibility violations. Further, \MotorEase analyzes \textit{both} visual UI screenshots and \texttt{\small uiautomator} metadata, whereas AidUI operates only upon visual UI screenshots -- with the use of multimodal data serving as a source of novelty for \MotorEase.

\section{Conclusion \& Future Work}
\label{sec:conclusion}

In this paper, we presented \MotorEase, an approach for detecting, classifying, reporting motor-impairment accessibility violations. We measured the performance, generalizability, and  applicability of \MotorEase to various open source applications. Our results indicate that \MotorEase is effective in practice and offers a novel approach for developers to identify accessibility issues affecting motor-impaired users. Future work will examine the potential to detect accessibility issues in web apps and conduct user~studies.

\section*{Acknowledgements}
\label{sec:acknowledgement}

This work is supported in part by NSF grants CCF-1955853 and CNS-2132285 and a gift from Dragon Testing Ltd. Any opinions, findings, and conclusions expressed herein are the authors' and do not necessarily reflect those of the sponsors. We would also like to express our sincere thanks to the authors of the Groundhog tool for their assistance in running the tool on the \MotorCheck benchmark.

\balance
\bibliographystyle{abbrv}
\bibliography{bibliography,ref}

\end{document}